\documentclass[%
 reprint,
 amsmath,amssymb,
prb,
]{revtex4-2}

\usepackage{graphicx}
\usepackage{dcolumn}
\usepackage{bm}
\usepackage{comment}

\newcommand*\diff{\mathop{}\!\mathrm{d}}

\begin{document}

\preprint{APS/123-QED}

\title{Relation between Photon Thermal Hall Effect and Persistent Heat Current in Nonreciprocal Radiative Heat Transfer}

\author{Cheng Guo}
 \affiliation{Department of Applied Physics, Stanford University, Stanford, California 94305, USA}

\author{Yu Guo}
\affiliation{%
 Ginzton Laboratory and Department of Electrical Engineering, Stanford University, Stanford, California 94305, USA
}

\author{Shanhui Fan}
\email{shanhui@stanford.edu}
\affiliation{%
 Ginzton Laboratory and Department of Electrical Engineering, Stanford University, Stanford, California 94305, USA
}

\date{\today}

\begin{abstract}
We study the photon thermal Hall effect and the persistent heat current in radiative heat transfer. We show that the photon thermal Hall effect is not a uniquely nonreciprocal effect; it can arise in some reciprocal systems with broken mirror symmetry. This is in contrast with the persistent heat current, which is a uniquely non-reciprocal effect that can not exist in any reciprocal system. Nevertheless, for a specific class of  systems with $C_4$ rotational symmetry, we note that the photon thermal Hall effect is uniquely nonreciprocal, and moreover there is a direct connection between the persistent heat current and the photon thermal Hall effect. In the near-equilibrium regime, the magnitude of the photon thermal Hall effect is proportional to the temperature derivative of the persistent heat current in such systems. Therefore, the persistent heat current as predicted for the equilibrium situation can be probed by the photon thermal Hall effect away from equilibrium.  
\end{abstract}

\maketitle


\section{\label{sec:intro} Introduction}

Thermal radiation is a ubiquitous physical phenomenon and its control  is important for both  science and engineering \cite{planck2013theory, More1971, rytov1989priniciples,  modest2013radiative,howell2015thermal,zhang2007nano, Joulain2005, Fan2017, Cuevas2018}. The vast majority of the literature on radiative heat transfer in general, and near-field heat transfer in particular, assumes materials that satisfy Lorentz reciprocity. On the other hand, recently there have been emerging interests exploring the unique aspects of thermal radiative heat transfer with the use of non-reciprocal materials \cite{Zhu2014a,Moncada-Villa2015,Ben-Abdallah2016,Zhu2016,Latella2017,Miller2017i,AbrahamEkeroth2018,Zhu2018,Zhao2019d}.

In near-field heat transfer, in particular, Zhu and Fan have pointed out the existence of the persistent current in an array of magneto-optical nanoparticles at equilibrium \cite{Zhu2016}. They have further noted that such a persistent current is a uniquely nonreciprocal effect -- such a current can not exist in any reciprocal system. Ben-Abdallah has noted the possibility of the photon thermal Hall effect, also in an array of magneto-optical nanoparticles, but  out of thermal equilibrium \cite{Ben-Abdallah2016}. It remains, however, an open question as to whether such a photon thermal Hall effect is a uniquely nonreciprocal effect, i.e., it can not exist in a reciprocal system. 

The contributions of the present paper are two-fold: (1) We show that the photon thermal Hall effect in general is not a uniquely nonreciprocal effect. It can also occur in a reciprocal system provided that the system breaks the relevant spatial symmetry. This is in contrast with the persistent heat current which requires nonreciprocity.  (2) For a specific class of systems with certain symmetry such as four-fold rotational symmetry, the photonic thermal Hall effect is a uniquely nonreciprocal effect. In such a case, there is  a connection between the persistent heat current and the photon thermal Hall effect. In the near-equilibrium case, the magnitude of the photon thermal Hall effect is directly proportional to the temperature derivative of the persistent heat current  in such  systems. Thus, in these specific systems, the photon thermal Hall effect in fact can be used to provide a direct experimental evidence of the persistent heat current. Our work provides a clarification of the role that nonreciprocity can play in near-field heat transfer, and may prove useful in the exploration of nonreciprocal heat transfer for device applications. 

The rest of this paper is organized as follows.
In Sec.~\ref{sec:background}  we briefly review the previous works on the photon thermal Hall effect and the persistent heat current in radiative heat transfer. 
In Sec.~\ref{sec:reciprocal} we numerically demonstrate that the photon thermal Hall effect can occur in reciprocal systems, thus does not require nonreciprocity. In Sec.~\ref{sec:connection} we analytically show that, in a system with four-fold rotational symmetry, there is a connection of the persistent heat current and the photon thermal Hall effect. We provide numerical evidence to support the analytic theory.  We conclude in  Sec.~\ref{sec:conclusion}. 
\section{\label{sec:background}Background}

\begin{figure}[htb]
\centering
\includegraphics[width=0.85\columnwidth]{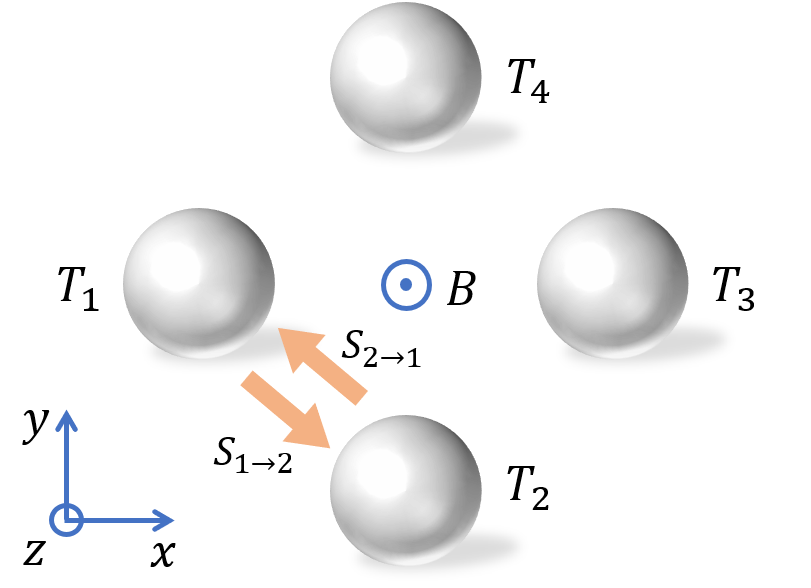}
\caption{\label{fig:geometry} Geometry of the system consisting of four magneto-optical spheres. The centers of the spheres are placed at the vertices of a square on the $x$-$y$ plane. A magnetic field is applied in the $z$ direction. 
The photon thermal Hall effect refers to a temperature gradient along the $y$ direction ($T_2-T_4$) due to the applied temperature gradient along the $x$ direction ($T_1-T_3$). The persistent heat current refers to a net heat flow between two bodies $($e.g.~$S_{1\rightarrow2}-S_{2\rightarrow1})$ at thermal equilibrium. For numerical demonstrations shown in the paper, these spheres are made of InSb. Each sphere has a radius of $100$ nm. The side length  of the square is $320$ nm.}
\end{figure}

We start by briefly reviewing the photon thermal Hall effect and the persistent heat current in radiative heat transfer. We consider the radiative heat transfer in the system as shown in Fig.~\ref{fig:geometry}. The system consists of four magneto-optical spheres. The centers of the spheres form a square on the $x$-$y$ plane. An external $B$ field is applied in the $z$ direction. In the presence of the magnetic field, the dielectric function of the  spheres becomes
\begin{equation}\label{eq:epsilon_general}
    \overline{\overline{\epsilon}}(B) = \begin{bmatrix}
    \epsilon_x & -i\epsilon'(B) & 0 \\
    i\epsilon'(B) & \epsilon_x & 0 \\
    0 & 0 & \epsilon_z
    \end{bmatrix},
\end{equation}
where $\epsilon'(B)=-\epsilon'(-B)$ such that $\overline{\overline{\epsilon}}(B)^T = \overline{\overline{\epsilon}}(-B)$ per Onsager reciprocal relations \cite{Onsager1931,LarseOnsager1931,Casimir1945, L.1980}.

For the system in Fig.~\ref{fig:geometry}, we first briefly review the persistent directional heat current at thermal equilibrium \cite{Zhu2016}. Suppose all the spheres have the same temperature. With an external field ($B\neq0$) that breaks reciprocity, the system in Fig.~\ref{fig:geometry} can exhibit a persistent heat current, as represented by the difference in the directional heat flow between bodies 1 and 2, $S_{1\rightarrow2}\neq S_{2\rightarrow1}$. Here $S_{i\rightarrow j}$ denotes the heat flow to body $j$ due to thermal excitation in body $i$. Such an effect is a uniquely non-reciprocal effect, as it has been proved that for a reciprocal system, there cannot be any net heat flow between bodies 1 and 2 ($S_{1\rightarrow2} =  S_{2\rightarrow1}$) if they have the same temperature, regardless of the temperature of other bodies \cite{Kruger2012,Zhu2016,Zhu2018}. 

We then briefly review the photon thermal hall effect \cite{Ben-Abdallah2016}. Suppose spheres 1 and 3 are held at two different temperatures $T_1$ and $T_3$ via contact with two heat baths. Due to the radiative heat exchange between the spheres, spheres 2 and 4 will reach temperatures $T_2$ and $T_4$ at steady state.  Without an external field ($B=0$), the system has mirror symmetry with respect to the  $x$-$z$ plane; therefore, $T_2 = T_4$. With an external field ($B\neq0$), such mirror symmetry is broken since $B$ is an axial vector whose sign flips under this mirror operation; therefore, as shown in  Ref.~\cite{Ben-Abdallah2016}, $T_2 \neq T_4$, i.e. a transverse 
temperature gradient along the $y$ direction develops due to the applied temperature gradient along the $x$ direction. This is the photon thermal Hall effect.

Even though the photon thermal Hall effect, in its original presentation \cite{Ben-Abdallah2016}, also uses magneto-optical particles, it is not clear whether such an effect fundamentally requires nonreciprocity. In the next section, we provide a numerical example of the photon thermal Hall effect to clarify this issue. 


\section{\label{sec:reciprocal} photon thermal Hall effect doesn't require nonreciprocity}

In this section, we show that the photon thermal Hall effect doesn't necessarily require nonreciprocity. We numerically demonstrate a reciprocal system that exhibits the photon thermal Hall effect, but no persistent heat current. 

We first derive the equations related to the photon thermal Hall effect that will be used for later calculation. We consider the radiative heat transfer in a general four-body system. The temperatures of bodies $1$ and $3$ are maintained at two different temperatures $T_1$ and $T_3$ by two heat baths. Bodies $2$ and $4$ are assumed to be isolated except through radiative heat contact with other bodies in the system and a common environment. At steady state, the temperatures of bodies $2$ and $4$ are determined by solving the equations requiring that the net heat flux into bodies $2$ and $4$ equals zero:
\begin{align}
    \label{eq:net_flux_general}
    0 = \int_0^\infty \diff{\omega}  \{&\sum_{j\neq i}[S_{j\rightarrow i }(\omega) - S_{i\rightarrow j }(\omega)] \notag \\
    &+ S_{env\rightarrow i}(\omega) -S_{i\rightarrow env}(\omega)\}, \quad i=2,4
\end{align}

In Eq.~(\ref{eq:net_flux_general}) 
\begin{align}\label{eq:S_F_general}
    S_{i\rightarrow j }(\omega) &= \frac{\Theta(\omega, T_i)}{2\pi} F_{i\rightarrow j}(\omega) \notag\\ 
    S_{i\rightarrow env }(\omega) &= \frac{\Theta(\omega, T_i)}{2\pi} F_{i\rightarrow env}(\omega) \notag\\ 
    S_{env \rightarrow j }(\omega) &= \frac{\Theta(\omega, T_{env})}{2\pi} F_{env\rightarrow j}(\omega)
\end{align}
where $\Theta(\omega, T) = \hbar \omega/[\operatorname{exp}(\hbar \omega/k_B T)-1]$, and $F_{a\rightarrow b}(\omega)$ denotes the temperature independent transmission coefficient from object $a$ to $b$ \cite{Zhu2018}.

We first provide a theoretical discussion showing that the photon thermal Hall effect can occur in reciprocal systems. To simplify the analytic treatment, we consider the linear-response regime, where the systems are near thermal equilibrium (i.e.~$T_j = T_{eq} + \Delta T_j, T_{env}=T_{eq}$, with $\Delta T_j\ll T_j$). Eq.~(\ref{eq:net_flux_general}) and Eq.~(\ref{eq:S_F_general}) then become:
\begin{align}
    \label{eq:net_flux_linear}
    0 = \sum_{j\neq i}[G_{j\rightarrow i }\Delta T_j - G_{i\rightarrow j }\Delta T_i] -G_{i\rightarrow env}\Delta T_i, \quad i=2,4,
\end{align}
where
\begin{align}
    \label{eq:thermal_conductance}
    G_{i\rightarrow j } \equiv \int_0^\infty \frac{\diff{\omega}}{2\pi} &\left.\frac{\partial \Theta(\omega, T)}{\partial T}\right\vert_{T=T_{eq}} F_{i\rightarrow j}(\omega),     \notag \\
    G_{i\rightarrow env } \equiv \int_0^\infty \frac{\diff{\omega}}{2\pi} &\left.\frac{\partial \Theta(\omega, T)}{\partial T}\right\vert_{T=T_{eq}} F_{i\rightarrow env}(\omega).     
\end{align}
are the thermal conductance from body $i$ to body $j$ and to the environment, respectively, at temperature $T_{eq}$. These thermal conductance in general are  non-negative since the system considered here is passive. 

Solving  Eq.~(\ref{eq:net_flux_linear}) we get

\begin{align}
    \label{eq:solution_linear}
    \Delta T_2 = \frac{1}{\Upsilon}\{&[G_{1\rightarrow2} (\sum_{j\neq4}G_{4\rightarrow j}+G_{4\rightarrow env})+G_{1\rightarrow4}G_{4\rightarrow2}]\Delta T_1 \notag\\
    + &[G_{3\rightarrow2} (\sum_{j\neq4}G_{4\rightarrow j}+G_{4\rightarrow env})+G_{3\rightarrow4}G_{4\rightarrow2}]\Delta T_3 \},   \notag \\
    \Delta T_4 = \frac{1}{\Upsilon}\{&[G_{1\rightarrow4} (\sum_{j\neq2}G_{2\rightarrow j}+G_{2\rightarrow env})+G_{1\rightarrow2}G_{2\rightarrow4}]\Delta T_1 \notag\\
    + &[G_{3\rightarrow4} (\sum_{j\neq2}G_{2\rightarrow j}+G_{2\rightarrow env})+G_{3\rightarrow2}G_{2\rightarrow4}]\Delta T_3 \},
\end{align}
where 
\begin{align}
    \Upsilon = (\sum_{j\neq2}G_{2\rightarrow j}+G_{2\rightarrow env})&(\sum_{j\neq4}G_{4\rightarrow j}+G_{4\rightarrow env}) \notag \\
    &-G_{2\rightarrow4}G_{4\rightarrow2} > 0.
\end{align}

The magnitude of the photon thermal Hall effect can be evaluated using the relative Hall temperature difference \cite{Ben-Abdallah2016}
\begin{equation}
\label{eq:relative_Hall}
    R = \frac{\Delta T_2 - \Delta T_4}{\Delta T_1 - \Delta T_3}
\end{equation}

Assuming the temperatures of thermostated body 1 and body 3 deviate anti-symmetrically from thermal equilibrium, i.e.~$\Delta T_1 = - \Delta T_3$,

\begin{align}
\label{eq:relative_Hall_sym}
    R = \frac{1}{2\Upsilon}[(G_{1\rightarrow 2}-G_{3\rightarrow2})(\sum_{j\neq 4}G_{4\rightarrow j}+G_{4\rightarrow env} - G_{2\rightarrow 4}) \notag\\
    + (G_{3\rightarrow 4}-G_{1\rightarrow4})(\sum_{j\neq 2}G_{2\rightarrow j}+G_{2\rightarrow env} - G_{4\rightarrow 2}) ].
\end{align}


Eq.~(\ref{eq:relative_Hall_sym}) is general for both reciprocal and nonreciprocal systems. If the system satisfies Lorentz reciprocity, $G_{i\rightarrow j} = G_{j\rightarrow i}$, Eq.~(\ref{eq:relative_Hall_sym}) simplifies to
\begin{align}\label{eq:R_reciprocity}
    &R = \frac{1}{2\Upsilon}[2(G_{1\rightarrow2} G_{3\rightarrow4}-G_{2\rightarrow3}G_{1\rightarrow4}) +\notag\\
    &(G_{1\rightarrow2}-G_{2\rightarrow3})G_{4\rightarrow env} +(G_{3\rightarrow4}-G_{1\rightarrow4})G_{2\rightarrow env}].
\end{align}

If one can further neglect the far field thermal exchange with the environment, Eq.~(\ref{eq:R_reciprocity}) becomes
\begin{align}\label{eq:R_reciprocity_noenv}
    R = \frac{1}{\Upsilon}(G_{1\rightarrow2} G_{3\rightarrow4}-G_{2\rightarrow3}G_{1\rightarrow4}),
\end{align}
which identifies with Eq.~(14) in Ref.~\cite{Ben-Abdallah2016}. Note  Eq.~(\ref{eq:R_reciprocity}) and Eq.~(\ref{eq:R_reciprocity_noenv}) only apply to reciprocal systems.

\begin{figure}[htbp]
\centering
\includegraphics[width=0.65\columnwidth]{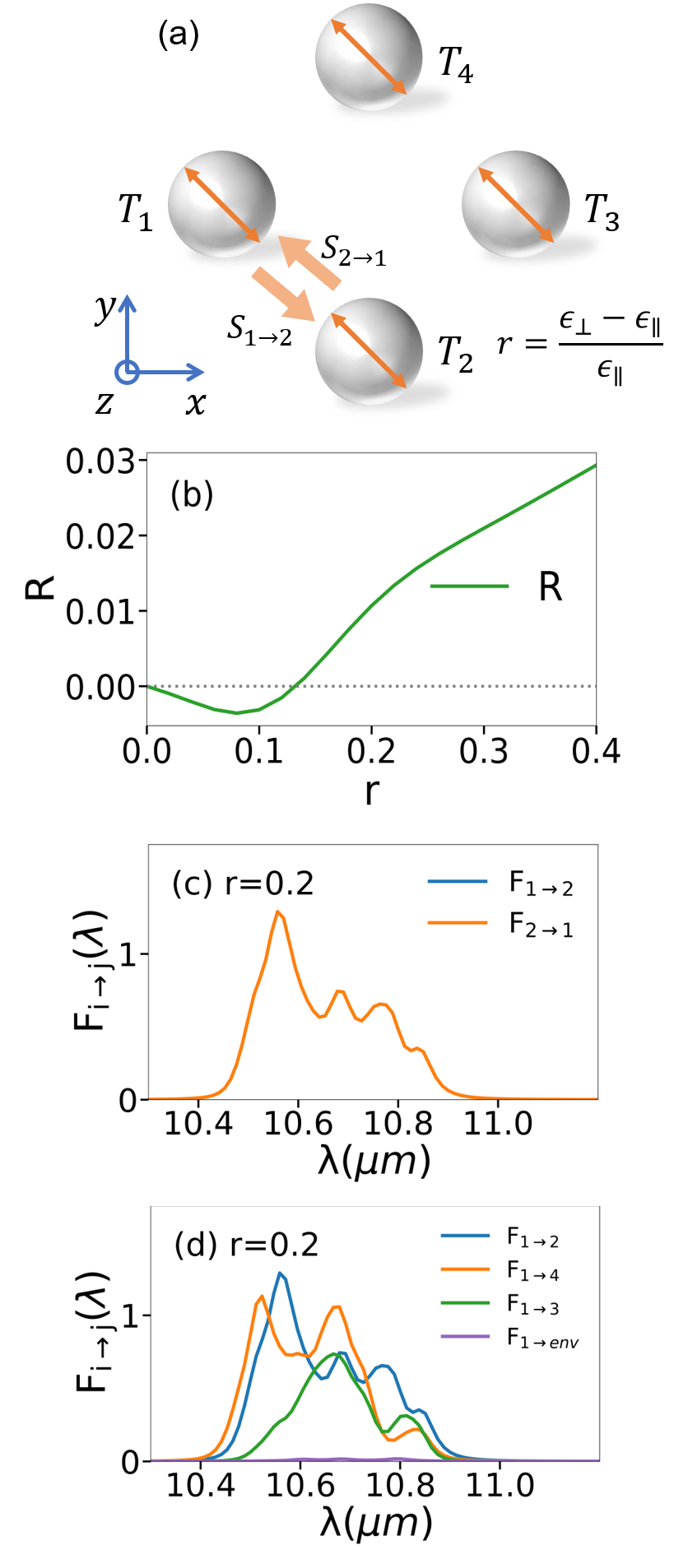}

\caption{\label{fig:strain} (a) The system consists of four InSb  spheres with a square geometry. Each sphere has a radius of $100$ nm. The side length  of the square is $320$ nm. A uniaxial stress is applied in the $(\hat{x}-\hat{y})$ direction, which induces uniaxial birefringence in InSb characterized by $r = (\epsilon_\bot-\epsilon_\parallel)/\epsilon_\parallel$. (b) $r$ dependence of the relative Hall temperature difference $R$. (c) The transmission coefficient spectra  $F_{1\rightarrow 2}(\lambda)$ and $F_{2\rightarrow 1}(\lambda)$. (d) The transmission coefficient spectra between spheres, and between sphere and the environment. Only four spectra are distinct due to the symmetry and reciprocity.}
\end{figure}


For a reciprocal system similar to what is shown in Fig.~\ref{fig:geometry} except without the external magnetic field, suppose the system has a mirror symmetry with particles $1$ and $3$ lying on the mirror plane, one sees the following relations: $G_{1\rightarrow 2} = G_{1\rightarrow 4}$, $G_{2\rightarrow 3} = G_{3\rightarrow 4}$ and $G_{2\rightarrow env} = G_{4\rightarrow env}$, and hence $R = 0$ from Eq.~(\ref{eq:R_reciprocity}). If the mirror symmetry is broken, the relations as mentioned above are no longer present, and hence the photon thermal Hall effect may occur in a reciprocal system provided that certain spatial symmetry is broken. Below, we provide a numerical example of the photon thermal Hall effect in a reciprocal system. 

We again consider a system consisting of four spheres with a square geometry as depicted in Fig.~\ref{fig:strain}(a). Instead of an external magnetic field, a uniaxial stress is applied to each sphere in the $(\hat{x}-\hat{y})$ direction. Such a stress causes otherwise optically isotropic spheres to exhibit uniaxial birefringence $\epsilon_\parallel\neq \epsilon_\bot$, where $\epsilon_\parallel$ and $\epsilon_\bot$ are the components of permittivity for light polarization parallel and perpendicular to the strain axis. The system with stress is reciprocal since the permittivity tensor is still symmetric.  


In the configuration used for calculation, each sphere has a radius of $100$ nm. The centers of the four spheres are placed on the vertices of a square. The side length of the square is $320$ nm. The spheres are made of $n$-doped InSb. The uniaxial stress is applied in the $(\hat{x}-\hat{y})$ direction, along the $[001]$ axis of InSb for all the spheres. In the presence of the stress, the dielectric function of InSb becomes \cite{Yu1971,Raynolds1995} \begin{equation}\label{eq:epsilon_stress}
    \overline{\overline{\epsilon}}  = \begin{bmatrix}
    \frac{\epsilon_\bot+\epsilon_\parallel}{2} & \frac{\epsilon_\bot-\epsilon_\parallel}{2} & 0 \\
    \frac{\epsilon_\bot-\epsilon_\parallel}{2} & \frac{\epsilon_\bot+\epsilon_\parallel}{2} & 0 \\
    0 & 0 & \epsilon_\bot
    \end{bmatrix}
    = \epsilon_\parallel \begin{bmatrix}
    1+\frac{r}{2} & \frac{r}{2} & 0 \\
    \frac{r}{2} & 1+\frac{r}{2} & 0 \\
    0 & 0 & 1+r
    \end{bmatrix}
    ,
\end{equation}
where $r=(\epsilon_\bot-\epsilon_\parallel)/\epsilon_\parallel$ characterizes the degree of the  strain-induced birefringence. For  calculation, we neglect the frequency dependence of $r$ since the relevant frequency range is relatively narrow, and we use
\begin{equation}\label{eq:epsilon_parallel}
\epsilon_\parallel = \epsilon_b - \frac{\omega_p^2}{\omega(\omega+i\Gamma)}.
\end{equation}
Here, the first term is the background permittivity as taken from Ref.~\cite{palik1998handbook}, which includes contributions from both interband transition and lattice vibration. The second term corresponds to the free-carrier contribution. $\Gamma$ is the free-carrier relaxation rate, and $\omega_p = \sqrt{n_e e^2/(m^*\epsilon_0)}$
is the plasma frequency. Following Ref.~\cite{Zhu2018}, we use a doping concentration $n_e =1.36\times 10^{19} \mathrm{cm^{−3}}$, for which the experimentally characterized relaxation rate \cite{Law2014} is   
 $ \Gamma = 10^{12} s^{-1}$ and the effective electron mass \cite{Law2014,Byszewski1963} is $m^* = 0.08m_e$. 

Fig.~\ref{fig:strain}(b) plots the relative Hall temperature difference $R$ calculated using Eq.~(\ref{eq:relative_Hall_sym}), which shows a nonmonotonic dependency on $r$. The nonzero $R$ indicates the existence of the photon thermal Hall effect when $r\neq 0$. This numerical example indeed demonstrates that the photon thermal Hall effect can exist in reciprocal systems. 

Using this system, we also provide a numerical illustration that the persistent heat current does not occur in reciprocal systems, including systems without mirror symmetries. Fig.~\ref{fig:strain}(c) plots the transmission coefficient spectra $F_{1\rightarrow 2}(\lambda)$ and $F_{2\rightarrow 1}(\lambda)$ for the system with $r=0.2$. We see that  $F_{1\rightarrow 2}(\lambda) = F_{2\rightarrow 1}(\lambda)$, which implies the heat transfer $S_{1\rightarrow2}= S_{2\rightarrow1}$ and hence the lack of the persistent current when the spheres $1$ and $2$ are at the same temperature (Eq.~(\ref{eq:S_F_general})).

Fig.~\ref{fig:strain}(d) plots the heat transfer spectra between the spheres, and the far-field radiation from the spheres to the environment. The system has reciprocity. In addition, the system has mirror symmetry around the planes normal to $\hat{x} + \hat{y}$ and $\hat{x} - \hat{y}$. Hence, we have  
\begin{align}\label{eq:symmetry_strain}
        &F_{1\rightarrow 2}(\omega) = F_{2\rightarrow 1}(\omega) =
        F_{3\rightarrow 4}(\omega) = F_{4\rightarrow 3}(\omega) , \notag\\
        &F_{1\rightarrow 4}(\omega) = F_{4\rightarrow 1}(\omega) =
        F_{2\rightarrow 3}(\omega) = F_{3\rightarrow 2}(\omega) , \notag\\
        &F_{1\rightarrow 3}(\omega) = F_{3\rightarrow 1}(\omega) = 
        F_{2\rightarrow 4}(\omega) = F_{4\rightarrow 2}(\omega) , \notag\\
        &F_{1\rightarrow env}(\omega) = F_{env\rightarrow 1}(\omega) =
        F_{2\rightarrow env}(\omega) = F_{env\rightarrow 2}(\omega) = \notag\\
        &F_{3\rightarrow env}(\omega) = F_{env\rightarrow 3}(\omega) =
        F_{4\rightarrow env}(\omega) = F_{env\rightarrow 4}(\omega),
\end{align}
Consequently, only four transmission coefficient spectra are distinct, as  plotted in Fig.~\ref{fig:strain}(d).

Fig.~2(d) elucidates the origin of the photon thermal Hall effect in such a system. With the constraints in Eq.~(\ref{eq:symmetry_strain}),   Eq.~(\ref{eq:R_reciprocity}) becomes
\begin{align}\label{eq:R_strain}
    R = \frac{1}{\Upsilon}(G_{1\rightarrow2}-G_{1\rightarrow4})(G_{1\rightarrow2} +G_{1\rightarrow4}+G_{2\rightarrow env}).
\end{align}
Thus, the nonzero $R$ results from the difference in $G_{1\rightarrow2}$ and $G_{1\rightarrow4}$ due to the mirror symmetry breaking induced by the stress.

\section{\label{sec:connection}Connection of the persistent heat current and the photon thermal Hall effect in magneto-optical particle systems}

In this section, we return to the system of magneto-optical nanoparticles under an external magnetic field as shown in Fig.~\ref{fig:geometry}. We show that in this specific system, which has $C_4$ symmetry, the photon thermal Hall effect is a uniquely nonreciprocal effect, and moreover there is a connection between this effect and the presence of persistent heat current. In the near-equilibrium case, the magnitude of the photon thermal Hall effect is directly proportional to the temperature derivative of the persistent heat current in such a system.

From the $C_4$ symmetry, the system in Fig.~\ref{fig:geometry} satisfies 
\begin{align}\label{eq:symmetry_magnetic}
        &F_{1\rightarrow 2}(\omega) = F_{2\rightarrow 3}(\omega) =
        F_{3\rightarrow 4}(\omega) = F_{4\rightarrow 1}(\omega) , \notag\\
        &F_{2\rightarrow 1}(\omega) = F_{3\rightarrow 2}(\omega) =
        F_{4\rightarrow 3}(\omega) = F_{1\rightarrow 4}(\omega) , \notag\\
        &F_{1\rightarrow 3}(\omega) = F_{2\rightarrow 4}(\omega) = 
        F_{3\rightarrow 1}(\omega) = F_{4\rightarrow 2}(\omega) , \notag\\
        &F_{1\rightarrow env}(\omega) = F_{2\rightarrow env}(\omega) =
        F_{3\rightarrow env}(\omega) = F_{4\rightarrow env}(\omega) = \notag\\
        &F_{env\rightarrow 1}(\omega) = F_{env\rightarrow 2}(\omega) =
        F_{env\rightarrow 3}(\omega) = F_{env\rightarrow 4}(\omega).
\end{align}
With the constraints in Eq.~(\ref{eq:symmetry_magnetic}), Eq.~(\ref{eq:relative_Hall_sym}) becomes 
\begin{align}\label{eq:R_2PT}
    R = \frac{1}{\Upsilon}(G_{1\rightarrow2}-G_{2\rightarrow1}) (G_{1\rightarrow2}+G_{2\rightarrow1}+G_{2\rightarrow env}).
\end{align}

Since $ (G_{1\rightarrow2}+G_{2\rightarrow1}+G_{2\rightarrow env})>0,$ and $ \Upsilon>0$, the sign of $R$ is determined by $(G_{1\rightarrow2}-G_{2\rightarrow1})$:
\begin{align}\label{eq:R_prop}
    R &\propto (G_{1\rightarrow2}-G_{2\rightarrow1}) \notag\\
    &= \int_0^\infty \frac{\diff{\omega}}{2\pi} \left.\frac{\partial \Theta(\omega, T)}{\partial T}\right\vert_{T=T_{eq}} [F_{1\rightarrow 2}(\omega)-F_{2\rightarrow 1}(\omega)]. 
\end{align}

On the other hand, the persistent heat current as represented by the net heat flow from body 1 to body 2 at thermal equilibrium is:
\begin{align}\label{eq:Delta_S_int}
    \Delta S_{1\rightarrow2} = \int_0^\infty \frac{\diff{\omega}}{2\pi} \Theta(\omega, T) [F_{1\rightarrow 2}(\omega)-F_{2\rightarrow 1}(\omega)], 
\end{align}
with its temperature derivative
\begin{align}\label{eq:dT_Delta_S_int}
    &\left.\frac{\diff{}}{\diff{T}} \Delta S_{1\rightarrow2}\right\vert_{T=T_{eq}}  \notag\\
    &=\int_0^\infty \frac{\diff{\omega}}{2\pi} \left.\frac{\partial \Theta(\omega, T)}{\partial T}\right\vert_{T=T_{eq}}[F_{1\rightarrow 2}(\omega)
    -F_{2\rightarrow 1}(\omega)]. 
\end{align}

Therefore, the relative Hall temperature difference is proportional to the the temperature derivative of the  persistent heat current:
\begin{equation}\label{eq:R_heat}
    R \propto (G_{1\rightarrow2}-G_{2\rightarrow1}) = \left.\frac{\diff{}}{\diff{T}} \Delta S_{1\rightarrow2}\right\vert_{T=T_{eq}}.
\end{equation}
Since the persistent heat current is a uniquely nonreciprocal effect in general, the derivation here also indicates that in this specific system the photon thermal Hall effect is a uniquely non-reciprocal effect. 
\begin{figure}[htbp]
\centering
\includegraphics[width=0.65\columnwidth]{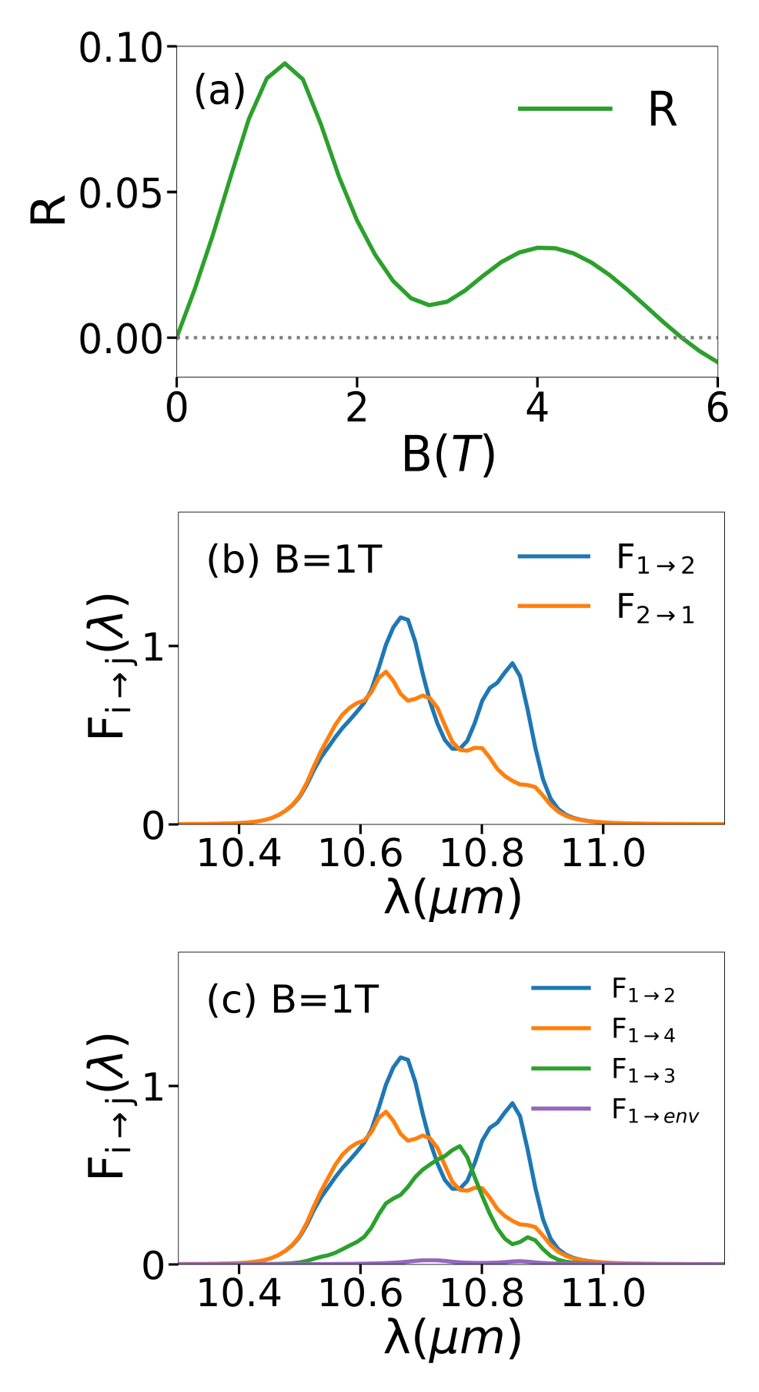}
\caption{\label{fig:magnetic} (a) $B$ dependence of the relative Hall temperature difference $R$. (b) The transmission coefficient spectra $F_{1\rightarrow 2}(\lambda)$ and $F_{2\rightarrow 1}(\lambda)$ when $B=1T$. (c) The transmission coefficient  spectra between spheres, and between spheres and environment. Only four  spectra are distinct due to the symmetry.}
\end{figure}

Now we numerically verify such a connection as indicated in Eq.~(\ref{eq:R_heat}). In the configuration used for calculation, each sphere has a radius of $100$ nm. The centers of the four spheres are placed on the vertices of a square. The side length of the square is $320$ nm. The spheres are made of $n$-doped InSb, which has a relative permittivity tensor
\begin{equation*}
    \overline{\overline{\epsilon}} = \epsilon_b \overline{\overline{I}} - \frac{\omega_p^2}{(\omega+i\Gamma)^2-\omega_c^2} 
    \begin{bmatrix}
    1 + i\frac{\Gamma}{\omega} & -i\frac{\omega_c}{\omega} & 0 \\
    i\frac{\omega_c}{\omega} & 1 + i\frac{\Gamma}{\omega} & 0 \\
    0 & 0 & \frac{(\omega+i\Gamma)^2-\omega_c^2}{\omega (\omega+i\Gamma)}
    \end{bmatrix}.
\end{equation*}
Here, the first term is the background permittivity as taken from Ref.~\cite{palik1998handbook}. The second term takes into account free-carrier contribution, which is sensitive to external magnetic field. $\Gamma$ is the free-carrier relaxation rate, $\omega_c = eB/m^*$ is the cyclotron frequency, and $\omega_p = \sqrt{n_e e^2/(m^*\epsilon_0)}$
is the plasma frequency. For calculation, we again use $n_e =1.36\times 10^{19} \mathrm{cm^{−3}}$,   
 $ \Gamma = 10^{12} s^{-1}$ and  $m^* = 0.08m_e$. 
 
Fig.~\ref{fig:magnetic}(b) plots the relative Hall temperature difference $R$ calculated using Eq.~(\ref{eq:relative_Hall_sym}), which shows a nonmonotonic dependency on $B$. The nonzero $R$ indicates the existence of the photon thermal Hall effect when $B\neq 0$. 

Fig.~\ref{fig:magnetic}(c) plots the transmission coefficient spectra  $F_{1\rightarrow 2}(\lambda)$ and $F_{2\rightarrow 1}(\lambda)$ when $B=1T$. We see the spectra are different, indicating a directional heat current in the counterclockwise direction.

Fig.~\ref{fig:magnetic}(d) plots the heat transfer spectra between the spheres, and the far-field radiation between the spheres and  the environment. 
Due to the constraints in Eq.~(\ref{eq:symmetry_magnetic}), only four transmission coefficient spectra are distinct, as plotted in Fig.~\ref{fig:magnetic}(d). 


\begin{figure}[htbp]
\centering
\includegraphics[width=0.98\columnwidth]{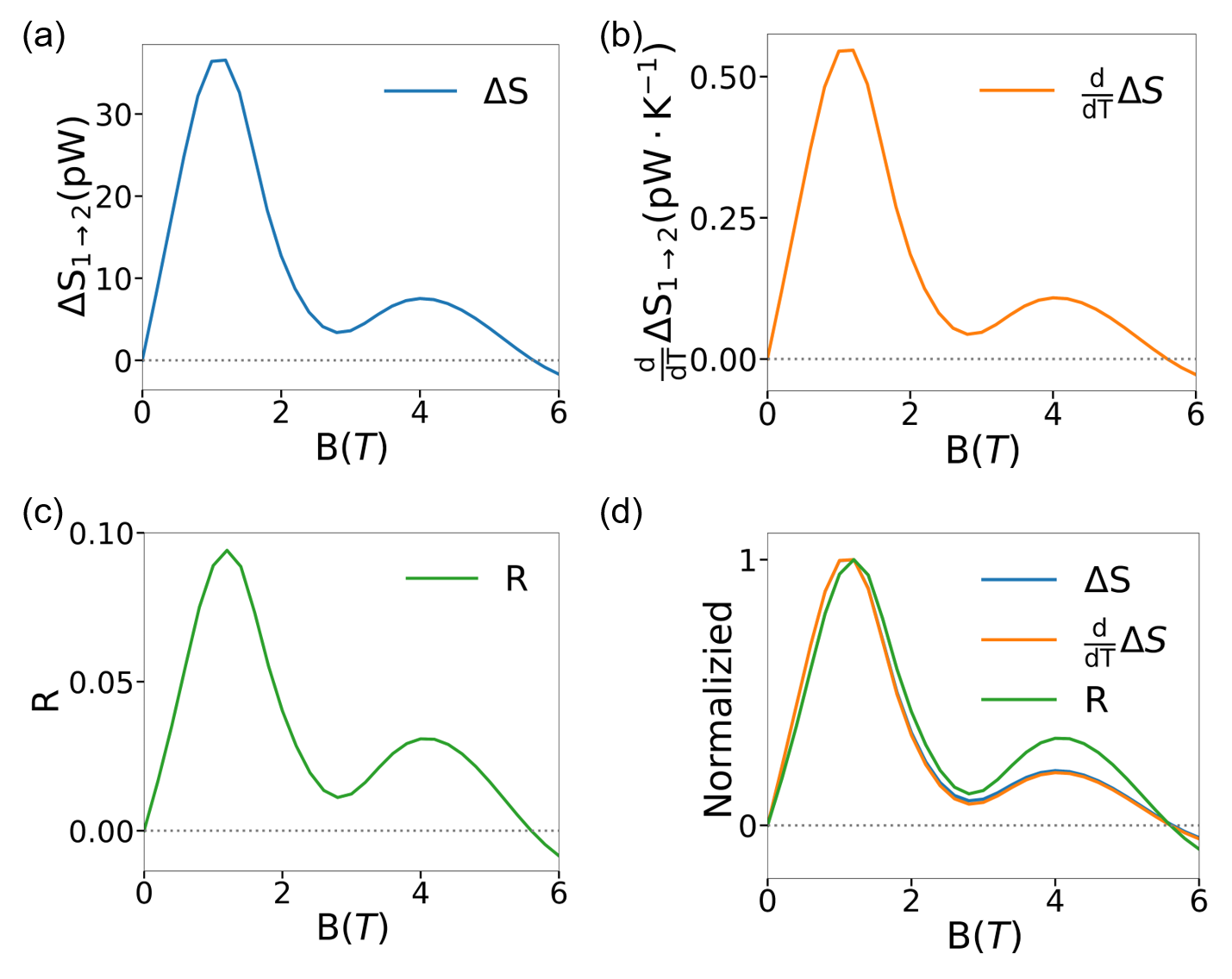}
\caption{\label{fig:connection} The magnetic field dependence of the persistent heat current and the photon thermal Hall effect for the system as shown in Fig.~\ref{fig:magnetic}(a). (a) the net heat flow from sphere 1 to sphere 2 at $\mathrm{T}=300\mathrm{K}$, and (b) its  temperature derivative. (c) The relative Hall temperature difference. (d) Normalized spectra for (a-c).}
\end{figure}

Finally, in Fig.~\ref{fig:connection} we illustrate the connection between the persistent heat current and the photon thermal Hall effect by their magnetic field dependence. 

Fig.~\ref{fig:connection}(a-c) plot the magnetic field dependence of $\Delta S_{1\rightarrow2}$, $\frac{\mathrm{d}}{\mathrm{d}T}\Delta S_{1\rightarrow2}$ and $R$ at $T=300K$, respectively. All these spectra show a similar nonmonotonic dependency on $B$. As a comparison, Fig.~\ref{fig:connection}(d) plots the three spectra normalized by their maximum values. We observe that the normalized $\Delta S$ and $\frac{\mathrm{d}}{\mathrm{d}T} \Delta S$ agree very well with each other. The normalized $R$ also follows the normalized $\frac{\mathrm{d}}{\mathrm{d}T} \Delta S$, especially for small magnetic field $B<3T$. The sign of $R$ agrees with that of $\frac{\mathrm{d}}{\mathrm{d}T} \Delta S$. There exists an optimal magnetic field strength which maximizes both the persistent heat current and the photon thermal Hall effect.  All these numerical observations are consistent with  the connection between these two effects (Eqs.~(\ref{eq:R_2PT}-\ref{eq:R_heat})). Such a connection thus suggests an experimental approach to quantitatively measure the persistent heat current by the photon thermal Hall effect. The deviation of the normalized $R$ and $\frac{\mathrm{d}}{\mathrm{d}T} \Delta S_{1\rightarrow 2}$ at large $B$ field is due to the small $B$  dependence of the factor  $\frac{1}{\Upsilon}(G_{1\rightarrow2}+G_{2\rightarrow1}+G_{2\rightarrow env})$ in Eq.~(\ref{eq:R_2PT}).

\section{\label{sec:conclusion}Conclusion}
In summary, we have studied the photon thermal Hall effect and the persistent heat current in radiative heat transfer.  We show that the photon thermal Hall effect is not a uniquely non-reciprocal effect; it can arise in some reciprocal systems. This is in contrast with the persistent heat current, which is a uniquely non-reciprocal effect that can not exist in any reciprocal system. Nevertheless, for a specific class of  systems with four-fold rotational symmetry, we note a direct connection between the persistent heat current and the photon thermal Hall effect. In the near-equilibrium regime, the magnitude of the photon thermal Hall effect is proportional to the temperature derivative of the persistent heat current in such systems. Therefore, the persistent heat current as predicted for the equilibrium situation can be probed by the photon thermal Hall effect away from equilibrium.

\section*{Acknowledgement}

C.G.~thanks Dr.~Linxiao Zhu for the numerical support, and Dr.~Bo Zhao for the stimulating discussions. This work is supported by a MURI grant from the U.~S.~Army Research Office (Grant No.~W911NF-19-1-0279), and by the Department of Energy “Photonics at
Thermodynamic Limits” Energy Frontier Research Center under Grant DE-SC0019140.

\end{document}